\documentclass{IEEEtran}
\usepackage{spconf,amsmath,graphicx}

\usepackage{amssymb,amsfonts,mathtools}
\usepackage{xcolor}
\usepackage{tikz}
\usepackage{pgfplots}
\DeclareUnicodeCharacter{2212}{−}
\usepgfplotslibrary{groupplots,dateplot}
\usetikzlibrary{patterns,shapes.arrows}
\pgfplotsset{compat=newest}
\usepackage{epsfig}
\usepackage{svg}
\usepackage[linesnumbered,ruled,vlined]{algorithm2e}
\usepackage{algorithmicx}
\usepackage{algpseudocode}
\usepackage[T1]{fontenc}
\usepackage{textcomp}
\usepackage{upquote}
\usepackage{multirow}
\usepackage{booktabs}
\usepackage{tablefootnote}
\usepackage{caption}
\usepackage{subfigure}
\usepackage{tabularx}
\usepackage{makecell}
\usepackage{adjustbox}
\usepackage{float}
\usepackage{framed}
\usepackage{ragged2e}
\usepackage{balance}
\usepackage{fancyhdr}
\usepackage{listings}
\usepackage{comment}
\usepackage{url}
\usepackage{lastpage}
\usepackage{academicons}
\usepackage{bm}

\usepackage{hyperref}
\usepackage{cite}

\newcommand{\supc}[1]{%
    (\tikz[baseline=(char.base)]{
        \node[shape=circle, fill=black, text=white, inner sep=0.5pt] (char) {\small #1};
    })
}

\newcommand{\papername}{HYPERDOA}
\title{\papername{}: Robust and Efficient DoA Estimation using Hyperdimensional Computing}

\name{{\it
Rajat Bhattacharjya\textsuperscript{1},\;
Woohyeok Park\textsuperscript{2},\;
Arnab Sarkar\textsuperscript{3},\
Hyunwoo Oh\textsuperscript{1},\;
Mohsen Imani\textsuperscript{1},\;
Nikil Dutt\textsuperscript{1}}}

\address{%
\textsuperscript{1}University of California, Irvine, USA; \
\textsuperscript{2}Kookmin University, Seoul, South Korea\\
\textsuperscript{3}Indian Institute of Technology, Kharagpur, India\\
\texttt{Corresponding author: rajatb1@uci.edu}
\thanks{Paper accepted at the IEEE International Conference on Acoustics, Speech, and Signal Processing (ICASSP) 2026. Authors' version posted for personal use and not for redistribution. The definitive version of the paper will appear in the proceedings of ICASSP 2026.}}

\begin{document}
\ninept
\maketitle


\begin{abstract}
Direction of Arrival (DoA) estimation techniques face a critical trade-off, as classical methods often lack accuracy in challenging, low signal-to-noise ratio (SNR) conditions, while modern deep learning approaches are too energy-intensive and opaque for resource-constrained, safety-critical systems. We introduce \texttt{\papername{}}, a novel estimator leveraging Hyperdimensional Computing (HDC). The framework introduces two distinct feature extraction strategies— \textit{Mean Spatial-Lag Autocorrelation} and \textit{Spatial Smoothing}—for its HDC pipeline, and then reframes DoA estimation as a pattern recognition problem. This approach leverages HDC's inherent robustness to noise and its transparent algebraic operations to bypass the expensive matrix decompositions and "black-box" nature of classical and deep learning methods, respectively. Our evaluation demonstrates that \texttt{\papername{}} achieves $\sim$35.39\% higher accuracy than state-of-the-art methods in low-SNR, coherent-source scenarios. Crucially, it also consumes $\sim$93\% less energy than competing neural baselines on an embedded NVIDIA Jetson Xavier NX platform. This dual advantage in accuracy and efficiency establishes \texttt{\papername{}} as a robust and viable solution for mission-critical applications on edge devices.
\end{abstract}


\section{Introduction}
\label{sec:intro}
In the context of array signal processing, \textit{Direction of Arrival} (DoA) estimation plays a critical role in localizing signal-emitting sources by determining the incident angles at which incoming signals impinge on an antenna array~\cite{deepmusic,deeprootmusic,dcd, doax, doax2, trans}. DoA estimation has numerous applications across domains such as autonomous vehicle localization~\cite{italy, access-av}, signal analysis in biomedical sensors~\cite{use2}, and seismic monitoring systems~\cite{seismic}. While a number of classical DoA estimation techniques exist---such as MUSIC~\cite{music}, Root-MUSIC~\cite{rootmusic}, and ESPRIT~\cite{esprit}---these methods often struggle to maintain high accuracy in challenging conditions, particularly under low signal-to-noise ratio (SNR) scenarios, with closely spaced or coherent sources, or in the presence of model mismatch. In addition to these accuracy limitations, their reliance on the process of subspace decomposition —invoking operations like eigenvalue decomposition (EVD) or singular value decomposition (SVD) — incurs high computational complexity, making them difficult to deploy efficiently in low-power embedded systems~\cite{music_lite}.


To overcome these limitations, recent research has turned to data-driven models like DeepMUSIC~\cite{deepmusic} and SubspaceNet~\cite{subspacenet}. While these deep learning-based approaches have shown improved robustness in low-SNR and multi-source scenarios, they introduce a critical trade-off. First, their "black-box" nature makes them difficult to analyze, verify, and trust—a significant drawback in safety-critical systems. Second, they often overlook system-level complexity, imposing substantial costs in terms of FLOPs, training overhead, and on-device power consumption, which hinders their viability for edge computing~\cite{access-av}.

Hence, to address both the reliability issues and system-level complexity of such data-driven DoA estimators, we propose the use of \textbf{Hyperdimensional Computing (HDC)}, a brain-inspired computational paradigm also known as Vector Symbolic Architecture (VSA)~\cite{hdc1,vsa}. HDC addresses the aforementioned challenges by design. Information is encoded in high-dimensional vectors (hypervectors), where the distributed representation provides intrinsic tolerance to noise, corruption, and hardware variability~\cite{hdc2}. Crucially, computation is performed using simple, massively parallel algebraic operations (e.g., bundling, binding, permutation~\cite{hdc1}). This unique combination of inherent robustness and computational efficiency makes HDC an ideal candidate for high-performance signal processing on low-power edge devices~\cite{biosig, flex}.

Building on this paradigm, we introduce \texttt{\papername{}}, a novel HDC-based DoA estimator that translates the theoretical advantages of HDC into a practical system. Our contributions are as follows:
\begin{enumerate}
    \item We design a complete HDC pipeline that reframes DoA estimation as a \textbf{pattern recognition problem}, using an associative memory for angle detection via similarity search. This approach entirely bypasses the need for expensive matrix factorizations during inference.
    \item We introduce two novel feature extraction variants for the HDC pipeline: \textbf{Mean Spatial-Lag Autocorrelation (Lag)} and \textbf{Spatial Smoothing} that help in providing higher accuracy than state-of-the-art (SOTA) DoA estimators, particularly in challenging low SNR and coherent-source scenarios by 35.39\%.
    \item We conduct a system-level energy evaluation on an embedded platform, the \textbf{NVIDIA Jetson Xavier NX}~\cite{jnx}, empirically confirming that \texttt{\papername{}} is significantly more energy-efficient than contemporary neural baselines by 92.93\%.
\end{enumerate}

\begin{figure*}
\centering
    \includegraphics[width=0.8\textwidth]{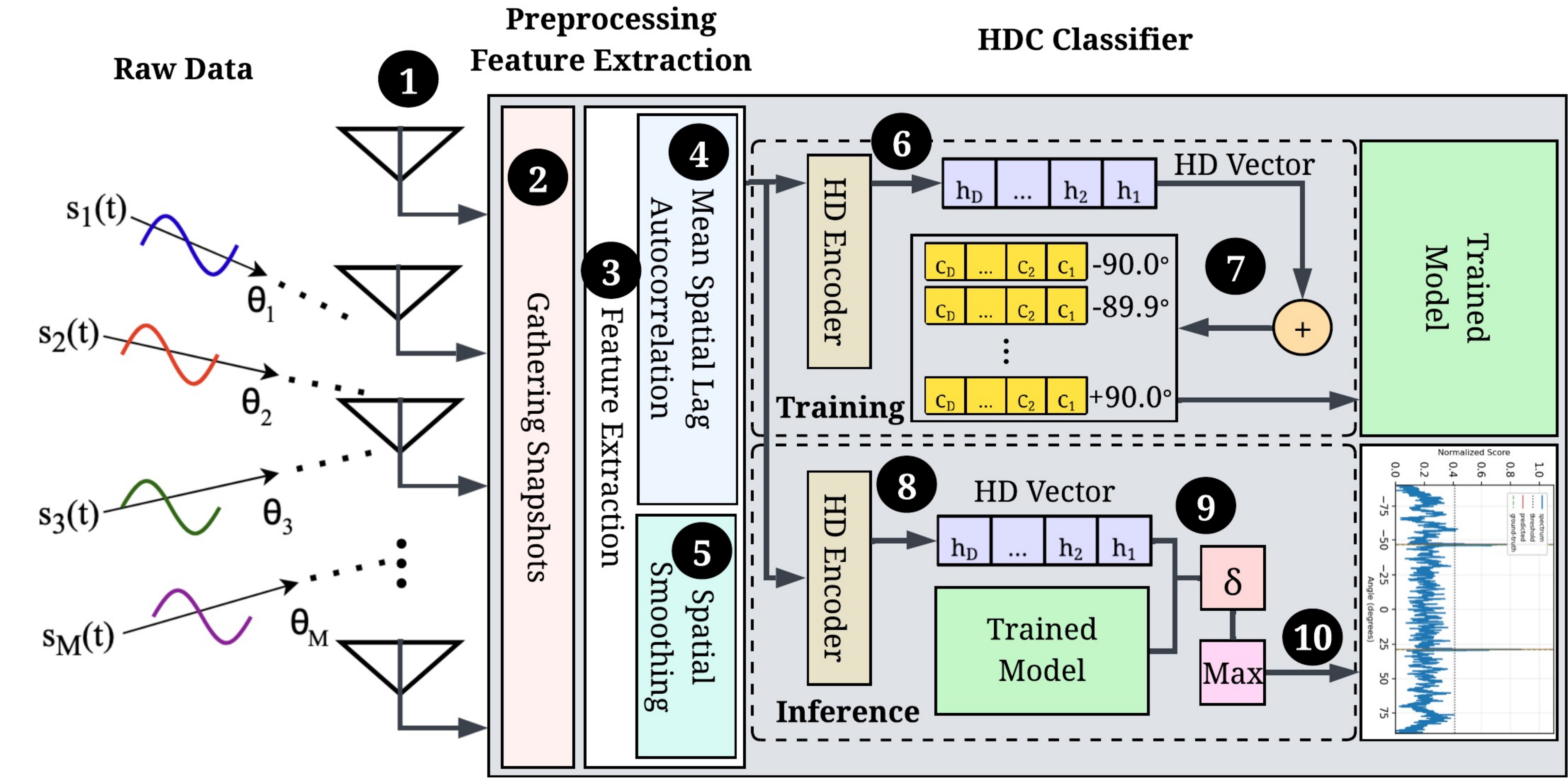}
    \caption{The \texttt{\papername{}} Architecture. A raw signal snapshot matrix $\mathbf{X}$ is first processed by the Feature Extraction module to generate a low-dimensional feature vector $\mathbf{f}$. This vector is then mapped by the HDC Encoder into a high-dimensional hypervector $\mathcal{H}_q$. The Associative Memory compares $\mathcal{H}_q$ against its stored prototype hypervectors (centroids) to \textbf{produce} an angular pseudo-spectrum of similarity scores. Finally, the Multi-Source Decoding stage identifies peaks in this spectrum to estimate the Direction of Arrival (DoA) for multiple sources.}
    \label{fig:arch}
\end{figure*}
\section{System Model}
\label{sec:sys}
We consider a uniform linear array (ULA) with $N$ antennas at $\lambda/2$ spacing, the signal received from $M$ narrowband sources over $T$ snapshots is modeled by the data matrix $\mathbf{X} \in \mathbb{C}^{N \times T}$:
\begin{align*}
    \mathbf{X} = \mathbf{A}(\boldsymbol{\theta})\mathbf{S} + \mathbf{V},
\end{align*}
where $\mathbf{A}(\boldsymbol{\theta}) \in \mathbb{C}^{N \times M}$ is the steering matrix for the source DoAs $\boldsymbol{\theta}$, $\mathbf{S} \in \mathbb{C}^{M \times T}$ contains the source signals, and $\mathbf{V} \in \mathbb{C}^{N \times T}$ is spatially white noise. The steering matrix is a concatenation of steering vectors, $\mathbf{A}(\boldsymbol{\theta}) = [\mathbf{a}(\theta_1), \dots, \mathbf{a}(\theta_M)]$, where for a ULA:
\begin{align*}
    \mathbf{a}(\theta) = [1, e^{-j\pi\sin\theta}, \dots, e^{-j\pi(N-1)\sin\theta}]^T.
\end{align*}
Many high-resolution algorithms operate on the spatial covariance matrix, $\mathbf{R_X}$, which is practically estimated using the sample covariance $\mathbf{\hat{R}_X} = \frac{1}{T}\mathbf{XX}^H$. The theoretical covariance matrix is $\mathbf{R_X} = \mathbf{A}(\boldsymbol{\theta})\mathbf{R_S}\mathbf{A}^H(\boldsymbol{\theta}) + \sigma_V^2\mathbf{I}_N$.

Assuming non-coherent sources, the eigendecomposition of $\mathbf{R_X}$ partitions the observation space into orthogonal signal and noise subspaces. Subspace methods like MUSIC exploit the orthogonality between true steering vectors and the noise subspace (spanned by $\mathbf{U}_N$), such that $\mathbf{a}^H(\theta_i)\mathbf{U}_N\mathbf{U}_N^H\mathbf{a}(\theta_i) = 0$.

However, the performance of these methods degrades significantly under low SNR, with few snapshots, or in the presence of coherent sources. They are also sensitive to model mismatches like array calibration errors. These limitations motivate the development of more robust and adaptive methods for DoA estimation such as \texttt{\papername{}}.

\section{\texttt{\papername{}} Architecture}
\label{sec:arch} 
Figure~\ref{fig:arch} outlines our \texttt{\papername{}} architecture
that reframes DoA estimation as a pattern recognition problem solvable with HDC, with a modular structure encompassing four main stages: (1) Feature Extraction, (2) HDC Encoding, (3) Associative Memory, and (4) Multi-Source Decoding.
\subsection{Feature Extraction}

The first stage of \texttt{\papername{}} distills the essential spatial information from the raw signal snapshot matrix into a compact feature vector. The process begins when raw signals impinge on the antenna array~\supc{1}, and their measurements are collected across multiple snapshots~\supc{2}. Given $M$ signal sources with directions of arrival $\{\theta_1, \ldots, \theta_M\}$, the antenna array captures the superposition of these signals across $N$ antennas over $T$ time snapshots, 
resulting in the raw signal matrix $\mathbf{X} \in \mathbb{C}^{N \times T}$.



Our HDC framework supports two feature extraction methods~\supc{3} as detailed below:\\
\noindent
\textbf{{\textit{\underline{Mean Spatial-Lag Autocorrelation (Lag):}}}} This feature~\supc{4} summarizes the spatial correlation structure of the wavefield.
First, we compute the sample spatial covariance matrix, a standard preprocessing step shared with classical methods like MUSIC~\cite{music}:
$
    \mathbf{\hat{R}_X} = \frac{1}{T}\mathbf{XX}^H.
$
From $\mathbf{\hat{R}_X}$, we extract the mean autocorrelation for each spatial lag $k \in \{0, \dots, N-1\}$ by averaging the elements along the $k$-th diagonal of the covariance matrix, producing a lag vector $\mathbf{r} \in \mathbb{C}^N$:
$$
    r_k = \frac{1}{N-k} \sum_{i=1}^{N-k} [\mathbf{\hat{R}_X}]_{i, i+k}.
$$
To reduce sensitivity to variations in signal power, the lag vector can be normalized by the magnitude of its zeroth element, $|r_0|$. The resulting complex-valued vector $\mathbf{r}$ is then separated into its real and imaginary components and concatenated to form a real-valued feature vector $\mathbf{f} \in \mathbb{R}^{2N}$. We apply z-score normalization~\cite{zscore} before HDC encoding. This method provides a low-dimensional representation of spatial correlations.\\
\noindent
\textbf{\textit{\underline{Spatial Smoothing:}}}
This feature~\supc{5} is commonly used to handle coherent signals, a case where subspace methods can degrade without smoothing. The core idea is to restore the rank of the covariance matrix by averaging over smaller, overlapping subarrays. The full antenna array of size $N$ is divided into $L$ overlapping subarrays, each of size $M_{\text{sub}} < N$. For each $j$-th subarray, we select the corresponding sensor data $\mathbf{X}_j \in \mathbb{C}^{M_{\text{sub}} \times T}$ and compute its sample covariance matrix: $\mathbf{R}_j = \frac{1}{T}\mathbf{X}_j\mathbf{X}_j^H$.

The final spatially smoothed covariance matrix is obtained by averaging these individual matrices:
$
    \mathbf{\hat{R}}_{SS} = \frac{1}{L} \sum_{j=1}^{L} \mathbf{R}_j.
$
This averaging process helps decorrelate coherent sources. The feature vector $\mathbf{f}$ is then constructed by vectorizing the upper triangular part of $\mathbf{\hat{R}}_{SS}$, separating the result into real and imaginary parts, and concatenating them. As with the previous method, z-score normalization is applied to the final feature vector.\\
\noindent
Hence, both feature extraction methods produce a real-valued feature vector $\mathbf{f}$ that serves as input to the HDC encoder.

\subsection{HDC Encoding}

We now move to HDC Encoding~\supc{6}, where the extracted feature vector $\mathbf{f}$ is mapped to a high-dimensional space using an HDC encoder. This mapping, $\mathcal{E}: \mathbb{R}^{2N} \to \mathbb{C}^D$, transforms the numeric features into distributed, holographic representations called hypervectors, where $D$ is the dimensionality ($D=10,000$ in our case). 

During training, \texttt{\papername{}} employs a Fractional Power Encoder based on Fourier Holographic Reduced Representations (FHRR)~\cite{mike} to encode the extracted feature vector. This encoder first assigns a unique, random base hypervector $\mathbf{B}_i \in \mathbb{C}^D$ to each of the $2N$ feature dimensions. These base hypervectors consist of complex numbers on the unit circle (i.e., random phases). The encoding process binds the feature values to these base hypervectors by treating each feature value $f_i$ as a phase rotation applied to its corresponding base vector $\mathbf{B}_i$. The final query hypervector $\mathcal{H}_q$ is generated by \textbf{binding}~\cite{hdc1} (element-wise multiplication in FHRR) the rotated base vectors: 
$
    \mathcal{H}_q = \bigotimes_{i=1}^{2N} \rho^{f_i}(\mathbf{B}_i), 
$
where $\otimes$ denotes element-wise multiplication (binding) and $\rho^{f_i}$ represents the phase rotation of each element in $\mathbf{B}_i$ by an angle proportional to the feature value $f_i$. This non-linear projection yields a hypervector for each input signal pattern, such that similar input features produce similar hypervectors in the high-dimensional space. 

\subsection{Associative Memory}

The core of the \texttt{\papername{}} model is an associative memory that stores a set of prototype hypervectors, or centroids, where each centroid $\mathcal{C}_{\theta}$ represents a discrete candidate angle $\theta$. The range of possible DoAs, $[\theta_{min}, \theta_{max}]$, is discretized with a predefined precision, and a unique centroid is learned for each discrete angle. 

During the training phase, the associative memory~\supc{7} is populated by adding query hypervectors to their corresponding angle centroids. This process uses an iterative learning rule fundamentally based OnlineHD~\cite{onlinehd}. However, OnlineHD is designed for single-label classification. In DoA estimation, a single input sample $\mathbf{X}$ can contain signals from multiple sources ($M > 1$), corresponding to a set of multiple true labels $\{\theta_1, \dots, \theta_M\}$. A direct application of the standard update rule—which reinforces the correct class and penalizes the single predicted (but incorrect) class—is problematic. This would cause conflicting updates, as a single sample would generate duplicate positive and negative updates among its multiple correct labels. 

To address this multi-label challenge, we adapt the learning rule as described below.
For a given input sample $\mathbf{X}$ with its corresponding query hypervector $\mathcal{H}_q$ and true labels $\{\theta_1, \dots, \theta_M\}$, our modified rule iterates through \emph{all} true labels. For each true label $\theta_i$ in the set, we apply \textbf{only a positive update}. The corresponding centroid $\mathcal{C}_{\theta_i}$ is moved closer to the query hypervector: 
\begin{align*} 
    \mathcal{C}_{\theta_i} \leftarrow \mathcal{C}_{\theta_i} + \eta \mathcal{H}_q, \quad \forall i \in \{1, \dots, M\} 
\end{align*} 
where $\eta$ is the learning rate. Crucially, we intentionally \textbf{omit the negative update} step for these multi-source samples. This adaptation allows a single query hypervector to be correctly associated with all its corresponding angle centroids simultaneously, without causing destructive interference. After training is complete, all centroids are normalized to unit length. This process creates a pattern memory where each centroid aggregates the hypervectors of all signals arriving from its corresponding angle, even when multiple signals are present in the same observation. 

During inference, the same FHRR HDC encoder~\supc{8} is used to transform the input features into a query hypervector. Then, we compute the dot product similarities~ \supc{9} between this query hypervector and all trained centroids. The vector of these similarities forms an angular pseudo-spectrum, analogous to the spatial spectrum in classical methods like MUSIC. 


\begin{figure*}
\centering
    \includegraphics[width=0.9\textwidth]{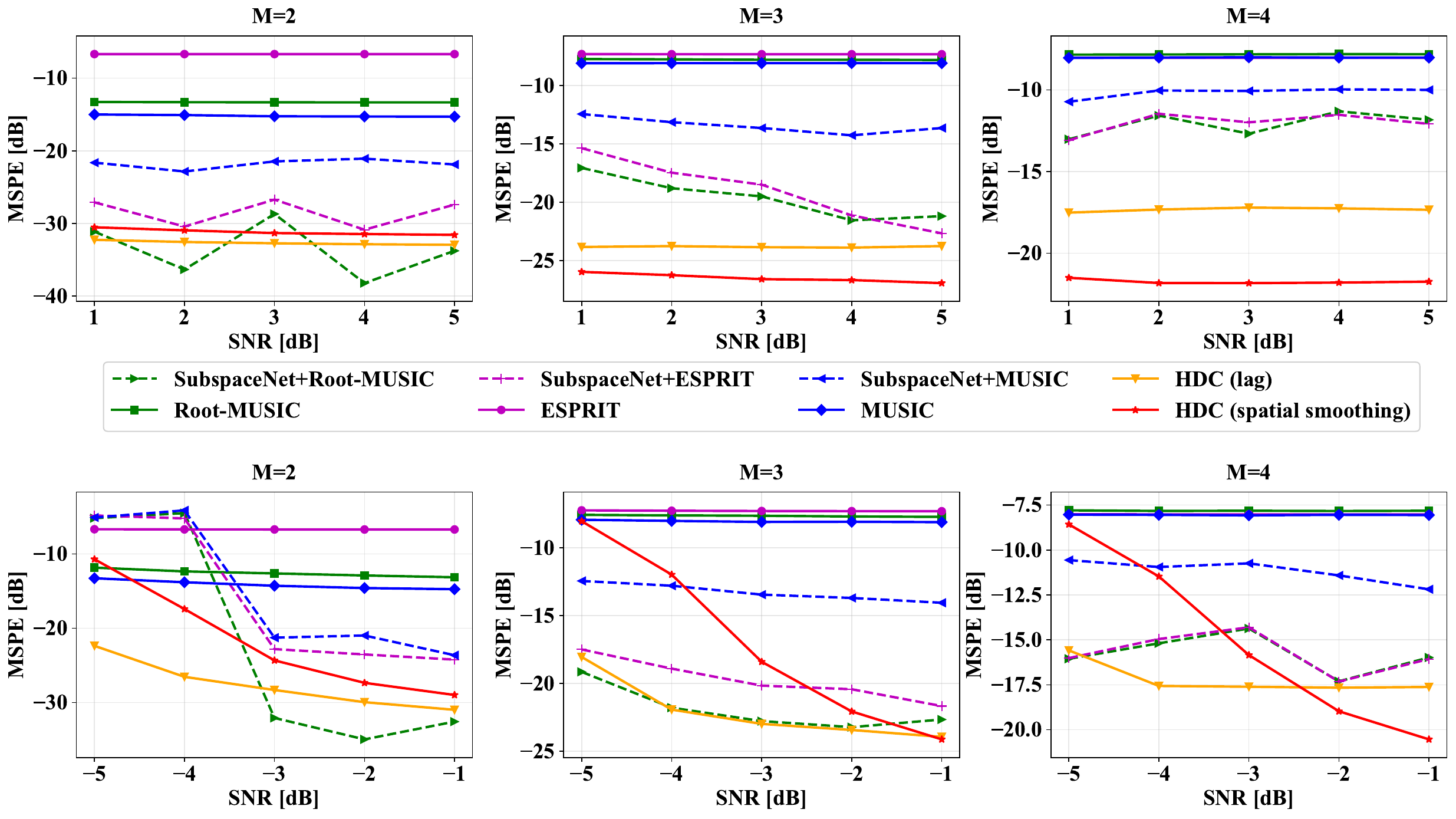}
    \caption{DoA Estimation MSPE ($T=100$, $N=8$, coherent sources) for SNR= [1,5] dB (upper row) and [-5,-1] dB (lower row).
}

    \label{fig:acc}
    \vspace{-3ex}
\end{figure*}
\subsection{Multi-Source Decoding}

Now, to finally estimate the DoAs of $M$ simultaneous sources~\supc{10}, \texttt{\papername{}} identifies the $M$ most prominent peaks in the angular pseudo-spectrum generated by the associative memory. This is accomplished using a non-maximum suppression algorithm~\cite{nmsx} that ensures the selected peaks are distinct and well-separated.

The decoding process operates in the following format: 
     (1) The algorithm first identifies the angle $\hat{\theta}_1$ corresponding to the global maximum of the similarity spectrum. (2) This angle is added to the set of estimated DoAs. (3) A suppression window is applied, centered at $\hat{\theta}_1$, where all similarity scores within a predefined minimum angular separation ($6^\circ$ in our case) are discarded. (4) The process is repeated on the remaining scores to find the next highest peak, $\hat{\theta}_2$, and so on, until $M$ sources have been identified.

This greedy peak selection resolves multiple sources by leveraging the full angular spectrum, providing a final set of estimated DoAs $\{\hat{\theta}_1, \dots, \hat{\theta}_M\}$. Unlike classical methods, this entire inference pipeline—from feature extraction to multi-source decoding—avoids computationally expensive matrix decompositions, which reduces computational burden and therefore benefits resource-constrained deployments.

\section{Experiments and Evaluation}
\begin{figure}
\centering
    \includegraphics[width=0.48\textwidth]{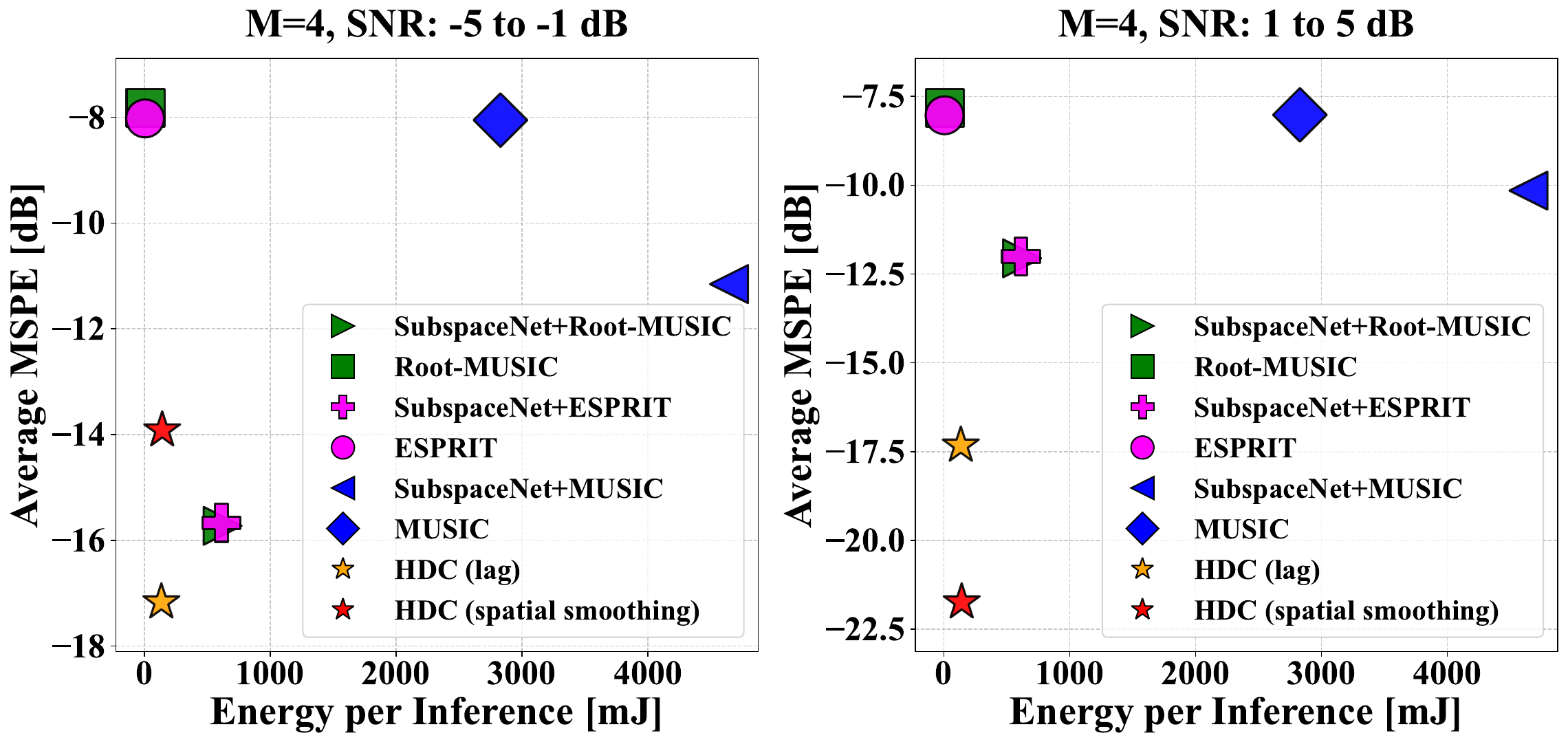}
    \caption{Accuracy vs. Energy Comparison ($M$=4, SNR=[-5,5] dB, $T$=100, $N$=8, coherent sources).
    }
    \label{fig:dse}
    \vspace{-3ex}
\end{figure}

\noindent
\textbf{\textit{\underline{Experimental Setup:}}}
 We evaluate \texttt{\papername{}} using a synthetic narrowband ULA model and report results under both non-coherent and coherent source scenarios while comparing it to SOTA methods\footnote{Source code: \url{https://github.com/pwh9882/HYPERDOA/}}. We use the following models:\\
\noindent
\vspace{1.8mm}
\textbf{\textit{\underline{Signal and array model:}}}
A half-wavelength spaced ULA is considered, with $N$ elements and $M$ sources impinging on the ULA. $T$ snapshots are collected. Source DoAs are sampled uniformly from $[-90^{\circ}, 90^{\circ}]$ with a minimum separation of $15^{\circ}$ enforced during data generation. We evaluate two cases: coherent and non-coherent. Our noise model is complex Gaussian.\\
\vspace{1.8mm}
\noindent
\textbf{\textit{\underline{Datasets and splits:}}}
For each configuration, we generate a training set of 45{,}000 samples and a test set sized at 5\% of the training set (2{,}250). Labels are the true DoAs in radians. \\
\vspace{1.8mm}
\noindent
\textbf{\textit{\underline{\texttt{\papername{}} parameters:}}} HDC encoder uses FHRR. Dimensionality ($D$) =10,000, angular grid resolution is $0.1^{\circ}$. For multi-source decoding, separation between peaks is  $6^{\circ}$.\\
\vspace{1.8mm}
\noindent
\textbf{\textit{\underline{Accuracy Analysis:}}}
\texttt{\papername{}} specifically aims for high accuracy in low SNR scenarios with multiple, coherent sources where SOTA models perform poorly. Our accuracy metric is mean squared periodic error (MSPE (dB))~\cite{subspacenet}. For the non-coherent source scenario with M=3, T=100, N=8, we see that HDC (lag) performs \textbf{6.29\% better} than SOTA methods within the SNR range of 1 to 5 dB, and in the lower SNR range of [-5,-1] dB, it provides \textbf{18.41\% better accuracy}. The results for coherent sources, shown in Figure~\ref{fig:acc}, demonstrate that for the SNR range of 1 to 5 dB with M=3 and 4, \texttt{\papername{}} performs the best. For instance, with M=3 in this range, HDC (lag) provides \textbf{45.38\% better accuracy} and HDC (spatial smoothing) provides \textbf{53.13\% better accuracy} than SOTA methods. At lower SNR levels from -1 to -5 dB, \texttt{\papername{}}'s performance remains the best among all candidates as the number of sources increases, thereby establishing HDC's inherent robustness. Overall, within the SNR range of [-5, 5] dB with M=3 and 4, \texttt{\papername{}} performs \textbf{35.39\%} better than SOTA methods.\\
\vspace{1.8mm}
\noindent
\textbf{\textit{\underline{Energy Consumption Study:}}}
We perform an energy consumption study of \texttt{\papername{}} and other SOTA methods on the NVIDIA Jetson Xavier NX embedded board, operating in its \texttt{MODE\_15W\_6CORE} power mode~\cite{power_m}. Over 2000 runs, we see that HDC (lag) consumes \textbf{135 mJ} of energy per inference, while HDC (spatial smoothing) consumes \textbf{142 mJ} per inference. In stark contrast, SubspaceNet+MUSIC consumes the most energy at \textbf{4645 mJ} per inference, which is \textbf{3253.8\%} higher than the average of HDC (lag) and HDC (spatial smoothing). Compared to all neural baselines (SubspaceNet+Root-MUSIC, SubspaceNet+ESPRIT, and SubspaceNet+MUSIC), \texttt{\papername{}} saves around \textbf{92.93\%} energy onboard, hence establishing itself as a core candidate for energy-constrained environments.\\
\noindent
\vspace{1.8mm}
\textbf{\textit{\underline{Accuracy-Energy Trade-offs:}}}
Having established the accuracy and energy consumption for different DoA estimators separately, we now study the accuracy-energy trade-offs, as shown in Figure~\ref{fig:dse}. The results highlight two key observations. First, in the low SNR range of -1 to -5 dB, HDC (lag) provides the highest accuracy while being the most energy-efficient compared to contemporary neural baselines. Second, for the SNR range of 1 to 5 dB, HDC (spatial smoothing) achieves the highest accuracy among all methods while substantially outperforming the neural baselines in energy efficiency. Overall, both HDC variants achieve top-tier accuracy while saving substantial energy onboard, thereby establishing \texttt{\papername{}}'s viability for resource-constrained, mission-critical environments.

\section{Conclusion and Future Work}
\label{sec:conclusion}
We introduce \texttt{\papername{}}, a new DoA estimator built on Hyperdimensional Computing (HDC). Unlike many existing methods that act as ``black boxes,'' consume large amounts of energy, and struggle in low SNR or coherent-source settings, \texttt{\papername{}} is designed to be both transparent and efficient.  It achieves this by exploiting HDC's inherent robustness and its transparent massively parallel algebraic operations. Furthermore, the framework introduces two distinct feature extraction strategies- \textit{Mean Spatial-Lag Autocorrelation} and \textit{Spatial Smoothing}—to specifically handle these challenging signal conditions. Our experiments show that \texttt{\papername{}} improves accuracy by $\sim$35.39\% compared to SOTA approaches in low SNR scenarios with coherent sources, and consumes $\sim$93\% less energy than neural baselines on an embedded NVIDIA Jetson Xavier NX device. These results support the use of \texttt{\papername{}} in resource-constrained, mission-critical scenarios. Looking ahead, we plan to test \texttt{\papername{}} in more challenging scenarios, study its resilience to array imperfections, and explore newer HDC techniques~\cite{ghrr, najafi}.

\section{Acknowledgements}
We thank Mike Heddes for his valuable input to this research.

\bibliographystyle{unsrt}

\end{document}